\newcommand*{\ATLASLATEXPATH}{latex/}
\documentclass[cernpreprint,texlive=2016,txfonts,UKenglish]{\ATLASLATEXPATH atlasdoc}
\usepackage[subfigure=true]{\ATLASLATEXPATH atlaspackage}
\usepackage{\ATLASLATEXPATH atlasbiblatex}
\usepackage{multirow}

\usepackage{\ATLASLATEXPATH atlascontribute}

\usepackage{\ATLASLATEXPATH atlasphysics}
\graphicspath{{logos/}{figures/}}

\def\wp{\ensuremath{W^\prime}}

\def\zp{\ensuremath{Z^\prime}}

\def\mt{$m_{\mathrm T}$}

\def\pythia{{\textsc{Pythia}}}

\def\geant{{\textsc{Geant}}}

\def\vrap{{\textsc{VRAP}}}

\def\sherpa{{\textsc{Sherpa}}}

\def\photos{{\textsc{Photos}}}

\def\powhegbox{{\textsc{Powheg-Box}}}

\def\mcsanc{{\textsc{Mcsanc}}}

\newcommand{\syspair}[2] { ${#1}$\%~(${#2}$\%) }

\def\wpssm{\ensuremath{W'_{\mathrm{SSM}}}} 
\def\xbr{\ensuremath{\sigma \times BR}}
\def\wpe{\ensuremath{W'\rightarrow e \nu}}
\def\wpmu{\ensuremath{W'\rightarrow \mu \nu}}

\def\wpl{\ensuremath{W'\rightarrow \ell \nu}}

\def\mwp{\ensuremath{m_{W'}}}

\AtlasTitle{\boldmath Search for a new heavy gauge-boson resonance decaying into a lepton and missing transverse momentum in 36 fb$^{-1}$ of $pp$ collisions at $\sqrt{s} = 13$~\tev\ with the ATLAS experiment }

\author{The ATLAS Collaboration}

\date{\today}

\PreprintIdNumber{CERN-EP-2017-082}

\AtlasJournal{EPJC}
\AtlasAbstract{%
The results of a search for new heavy \wp\ bosons decaying to an electron or muon and a neutrino using proton--proton collision data at a centre-of-mass energy of $\sqrt{s}~=~13$~\tev\ are presented. The dataset was collected in 2015 and 2016 by the ATLAS experiment at the Large Hadron Collider and corresponds to an integrated luminosity of $36.1$~\ifb.
As no excess of events above the
Standard Model prediction is observed, the results are used to set upper limits on the \wp\ boson cross-section times branching 
ratio to an electron or muon and a neutrino as a function of the \wp\ mass.
Assuming a \wp\ boson with the same couplings as the Standard Model $W$ boson, \wp\ masses below $5.1$~\tev\ are excluded at the 95\% confidence level. 
}

\hypersetup{pdftitle={ATLAS draft},pdfauthor={The ATLAS Collaboration}}

\begin{document}

\maketitle

%\tableofcontents

% List of contributors - print here or after the Bibliography.
%\PrintAtlasContribute{0.30}
%\clearpage

%-------------------------------------------------------------------------------
\section{Introduction}
\label{sec:intro}
%-------------------------------------------------------------------------------
Extensions to the Standard Model (SM) may include heavy gauge bosons that could be discovered at the Large Hadron Collider (LHC) \cite{LHC}. For example, they are predicted in left-right symmetric models \cite{Mohapatra:1974hk,Senjanovic:1975rk}
or the little Higgs model \cite{ArkaniHamed:2002qy}. Conceptually, these particles are heavier versions of the SM $W$ and $Z$ bosons and are generically referred to as \wp\ and \zp\ bosons. The Sequential Standard Model (SSM) \cite{ssm} posits a \wpssm\ boson with couplings to fermions that are identical to those of the SM $W$ boson. This model is a good benchmark as the results can be interpreted in the context of other models of new physics, and is useful for comparing the sensitivity of different experiments.

This paper presents a search for a \wp\ boson conducted in the $\wp \to \ell \nu$ channel. In the following, the term {\textit {lepton}} ($\ell$) is used to refer to an electron or a muon.
The analysis uses events with a high transverse momentum\footnote{ATLAS uses a right-handed coordinate system with its origin at the nominal interaction point (IP) in the centre of the detector and the $z$-axis along the beam pipe. The $x$-axis points from the IP to the centre of the LHC ring, and the $y$-axis points upward. Cylindrical coordinates $(r,\phi)$ are used in the transverse plane, $\phi$ being the azimuthal angle around the $z$-axis. The pseudorapidity is defined in terms of the polar angle $\theta$ as $\eta=-\ln\tan(\theta/2)$. Transverse momentum (\pt) is defined relative to the beam axis and is calculated as $\pt = p \sin(\theta)$ where $p$ is the momentum.} (\pT) lepton and significant missing transverse momentum \met, which is the measure to infer the presence of the neutrino in the event as it escapes direct detection. It is based on 36.1 \ifb\ of $pp$ collision data collected with the ATLAS detector in 2015 and 2016 at a centre-of-mass energy of $\sqrt{s} = 13$~\tev. 
The results are interpreted in the context of the SSM. 
The signal discriminant is the transverse mass, which is defined as $\smash[b]{m_{\mathrm T}} = \sqrt{\smash[b]{2\pT\met(1-\cos \phi_{\ell\nu})}}$, where $\smash[b]{\phi_{\ell\nu}}$ is the azimuthal angle between the lepton \pt\ and the \met{} directions in the transverse plane.

The most stringent limits on the mass of a \wpssm\ boson to date come from the searches in the $\wp \to e \nu$ and $\wp \to \mu \nu$ channels by the ATLAS and CMS collaborations using data taken at $\sqrt{s} = 13$~\tev\ in 2015. The ATLAS analysis was based on data corresponding to an integrated luminosity of 3.2~\ifb\ 
and sets a 95\% confidence level (CL) lower limit
on the \wpssm\ mass of 4.07~\TeV~\cite{atlas_13tev_pub}. The CMS Collaboration used 2.4~\ifb\ of data and excludes \wpssm\ masses below 4.1~\TeV\ at 95\%~CL~\cite{cms_13tev_pub}.
The sensitivity of the search presented here is significantly improved compared to these earlier searches due to the larger dataset.

%-------------------------------------------------------------------------------
\section{ATLAS detector}
\label{sec:detector}
%-------------------------------------------------------------------------------
The ATLAS experiment \cite{atlas:detector} at the LHC is a multipurpose particle detector with a forward-backward symmetric cylindrical geometry and a near $4 \pi$ 
coverage in solid angle. It consists of an inner
detector (ID) for tracking surrounded by a thin superconducting solenoid providing a \SI{2}{\tesla} axial magnetic field, electromagnetic (EM) and hadronic calorimeters, and a muon spectrometer (MS). The ID covers the pseudorapidity range $|\eta|<2.5$. It consists of a silicon pixel detector including an additional inner layer located at a radius of 3.2 cm since 2015 \cite{ibl:detector}, followed by silicon microstrip and transition radiation
tracking detectors. 
Lead/liquid-argon (LAr) sampling calorimeters provide EM energy measurements with high granularity. A hadronic (steel/scintillator-tile) calorimeter covers the central pseudorapidity range ($|\eta|<1.7$). The end-cap and forward regions are instrumented with LAr calorimeters for both the EM
and hadronic energy measurements up to $|\eta|=4.9$. The muon spectrometer surrounds the calorimeters and is based on three large air-core toroidal superconducting magnets with eight coils each. 
The field integral experienced by tracks in the toroidal field ranges between \num{2.0} and \SI{6.0}{\tesla\metre} for most pseudorapidities.
The MS includes a system of precision tracking chambers, over $|\eta| < 2.7$, and fast detectors for triggering, over $|\eta| < 2.4$. A two-level trigger system is used to select events \cite{atlas:trigger}. The first-level trigger is implemented in hardware and uses a subset of the detector information. This is followed by a software-based trigger system that reduces the accepted event rate to less than \SI{1}{\kilo\hertz}.

%-------------------------------------------------------------------------------
\section{Analysis strategy and modelling of signal and background processes}
\label{sec:MC}
%-------------------------------------------------------------------------------
A \wp\ signal would appear as an excess of events above the SM background at high \mt. This background mainly arises from processes with at least one prompt final-state lepton, with the largest source being the charged-current Drell--Yan (DY) $W$ boson production, where the $W$ boson decays into an electron or muon and a neutrino.
The second largest source is top-quark pair (\ttbar) and single-top-quark production, denoted in the following as
``top-quark background''. Other non-negligible contributions are from the neutral-current DY ($Z/\gamma^*$) process, diboson production, as well as from events  in which one final-state jet or photon satisfies the lepton selection criteria. The latter background component,
referred to in the following as the multijet background, receives contributions from
multijet, heavy-flavour quark and $\gamma$ + jet production. The multijet background is determined using a data-driven method, while the other backgrounds are modelled by Monte Carlo (MC) simulations.

The backgrounds from $W \rightarrow \ell \nu$, $Z/\gamma^*\rightarrow \ell \ell$, $W \rightarrow \tau \nu$, and $Z/\gamma^*\rightarrow \tau \tau$ 
were simulated using the \powhegbox\ v2~\cite{Alioli:2010xd} matrix-element calculation up to next-to-leading order (NLO) in perturbative quantum chromodynamics (pQCD), interfaced to the \pythia~8.186~\cite{pythia8} parton shower model and using the CT10 parton distribution function (PDF) set~\cite{CT10}. The final-state photon radiation (QED FSR) was modelled by the \photos~\cite{fsr_ref} MC simulation. 
The samples are normalised as a function of the boson invariant mass to a next-to-next-to-leading order (NNLO) pQCD calculation using the numerical programme \vrap\, which is based on Ref.~\cite{vrap} and the CT14NNLO PDF set~\cite{Dulat:2015mca}. Compared to the NLO prediction using CT10, the NNLO prediction using CT14 gives a higher cross-section by about 5\% at a boson invariant mass of 1 \tev\ and 
10\% at 5 \tev.
In addition to the modelling of QED FSR, a fixed-order electroweak (EW) correction to NLO is 
calculated as a function of the boson mass with the \mcsanc~\cite{Bardin:2012jk,Bondarenko:2013nu} event generator at leading order (LO) in pQCD. 
This correction is added to the NNLO QCD cross-section prediction in the so-called additive approach (see Section~\ref{sec:sysTheory}) because of a lack of calculations of mixed QCD and EW terms,
and lowers the predicted cross-section by an increasing amount as function of the mass, reaching about 10\% at 1 \tev\ and 20\% at 5 \tev.
The $W \rightarrow \ell \nu$ and $Z/\gamma^*\rightarrow \ell \ell$ events were simulated as multiple samples covering different ranges of the boson invariant mass. This ensures that a large number of MC events is available across the entire \mt\ region probed in this analysis.

The background from $t\bar{t}$ production was generated using \powhegbox\ v2, with parton showering and hadronisation modelled by \pythia~6.428~\cite{Sjostrand:2006za}, using the CT10 PDF set. The $t\bar{t}$ cross-section is normalised to $\sigma_{t\bar{t}}= 832$ pb 
as calculated with the Top++2.0 program at NNLO in pQCD, including soft-gluon resummation 
to next-to-next-to-leading logarithmic accuracy (see Ref.~\cite{Czakon:2011xx} and references therein). The top-quark mass is set to 172.5 \gev. 
The single-top-quark production in the $Wt$ channel and EW $t$-channel was simulated using the same event generators and PDF sets as for the $t\bar{t}$ process, with the exception that the \powhegbox\ v1 program was used for producing events in the $t$-channel. 
Diboson events were simulated with the \sherpa~2.1.1 \cite{Gleisberg:2008ta} event generator using the CT10 PDF set. 
As the simulated top-quark and diboson samples are statistically limited at large \mt, the expected number of events from each of these backgrounds is extrapolated into the high-\mt\ region. This is achieved by fitting the lower part of the \mt\ distributions to functions of the form $F (x) = a \, x^{b+c \log x}$ and $F (x) = d/ (x + e)^g$, where $x = m_{\mathrm T}/\sqrt{s}$, and using the fitted function to predict the background at higher \mt. Various fit ranges are used, which typically start between $140$ and $360$~\GeV\ and extend up to $500$ to $1300$~\GeV. The fits with the best $\chi^2/{\textrm {d.o.f.}}$ are used for the extrapolation and the results of these fits are used in the high-\mt\ tail. 

The multijet background is estimated from data using the same data-driven {\textit{matrix}} method as used in the previous ATLAS analysis~\cite{atlas_13tev_pub}. 
The first step of the matrix method is to calculate the fraction $f$ of lepton candidates that pass the nominal lepton identification and isolation requirements
({\textit{tight}}), 
with respect to a sample of {\textit{loose}} lepton candidates in a background-enriched sample.
These loosely selected candidates satisfy only a subset of the nominal criteria, which is
stricter than the trigger requirements imposed. 
Potential contamination of prompt final-state leptons in the background-enriched sample
is accounted for using MC simulation. In addition, the fraction $r$ of
real leptons in the sample
of {\textit{loose}} candidates satisfying the nominal requirements is used. 
This fraction is computed from MC simulation.
The number of jets and photons misidentified as leptons ($N_{\mathrm T}^{\mathrm{multijet}}$) in the total number of candidates passing the signal selection ($N_{\mathrm T}$) is
\begin{linenomath} 
\begin{equation}
\label{bkg_true_quantities}
N_{\mathrm T}^{\mathrm{multijet} }  = f \, N_{\mathrm F} = \frac{f}{r-f} \Big(\, r\,  (N_{\mathrm L} + N_{\mathrm T}) - N_{\mathrm T} \Big),
\end{equation}
\end{linenomath}
where $N_\mathrm{F}$ is the number of fake leptons and $N_\mathrm{L}$ corresponds to leptons that pass the {\textit{loose}} requirements but fail the nominal requirements. 
As this background estimate is statistically limited at large \mt, the expected number of events is extrapolated into the high-\mt\ region using a method similar to that for the diboson and top-quark backgrounds. 
 
The SSM signal $W^\prime \rightarrow e \nu$ and $W^\prime \rightarrow \mu \nu$ samples were generated at LO in QCD using the \pythia~8.183 event generator and the NNPDF2.3 LO PDF set~\cite{Ball:2012cx}. As assumed in the SSM, the couplings to fermions are equal to those of the SM $W$ boson. The \wp\ boson is assumed not to couple to the SM $W$ and $Z$ bosons and interference between the \wp\ and the SM $W$ boson production amplitudes is neglected. The decay $W^\prime \rightarrow \tau \nu$, where the $\tau$ subsequently decays leptonically, is not treated as part of the signal as this contribution was quantified previously and found to give a negligible contribution to the sensitivity~\cite{atlas_7tev_pub}. Mass-dependent correction factors are applied to normalise the samples to the same mass-dependent NNLO pQCD calculation as used for the $W$ background.
Compared to the LO prediction using NNPDF2.3 LO, the corrections increase the cross-section by about 40\% around a boson invariant mass of
1--2 \tev, and by about 10\% at 5 \tev.
Further EW corrections beyond QED FSR are not considered for the signal.
The resulting cross-sections times branching ratio for \wpssm\ masses of 3, 4 and 5~\tev\ are 15.3, 2.25 and 0.51~fb, respectively.
For these \wp\ masses the branching ratio to each lepton generation from \pythia\ is 8.2\%. 

The MC samples were processed through a simulation of the detector geometry and response \cite{SOFT-2010-01} using the \geant4~\cite{geant} framework. The software used for the reconstruction is the same for both simulated and real data. 
The average number of pile-up interactions (additional $pp$ collisions in the same or a nearby bunch crossing) observed in the data is about 23. The effect of pile-up is modelled by overlaying simulated inelastic $pp$ collision events selected using very loose trigger requirements (``minimum bias''). All MC samples are reweighted so that the distribution of the number of collisions per bunch crossing matches the data.
Correction factors to account for differences observed in the detector response between data and simulation are applied to the lepton trigger, reconstruction, identification \cite{elepaper, muonpubnote} 
and isolation efficiencies as well as the lepton energy/momentum resolutions and  
scales \cite{elecalibpaper, muonpubnote}.

%-------------------------------------------------------------------------------
\section{Event reconstruction}
\label{sec:selection}
%-------------------------------------------------------------------------------
The analysis makes use of electrons, muons, and missing transverse momentum, the reconstruction and identification of which are explained in the following.

Electrons are reconstructed from ID tracks that are matched to energy clusters in the electromagnetic calorimeter obtained using a sliding-window algorithm in the range $|\eta| < 2.47$. Candidates in the transition region between different electromagnetic calorimeter components, $1.37 < |\eta| < 1.52$, are rejected. Electrons must satisfy identification criteria based on measurements of calorimeter shower shapes and measurements of track properties from the ID combined in a likelihood discriminant. Depending on the desired level of background rejection, {\textit{loose}}, {\textit{medium}} and {\textit{tight}} working points are defined. Full details of the electron reconstruction, identification and selection working points can be found in Ref.~\cite{elepaper}.  

Muon candidates are identified from MS tracks that match tracks in the ID, with $|\eta|<2.5$ ~\cite{muonpubnote}. 
These muons are required to pass a track quality selection based on the number of hits in the ID. They are rejected if the absolute value of the difference between their charge-to-momentum ratios
measured in the ID and MS divided by the sum in quadrature of the
corresponding uncertainties is large.
To ensure optimal muon resolution at high \pt, additional requirements are imposed on the quality of the MS track. 
The track is required to have at least three hits in each of the three separate layers of MS chambers. Furthermore,  
to avoid \pt\ mismeasurements, muons are removed if they cross either poorly aligned MS chambers, or regions in which the ID and the MS are not well aligned relative to one another.

The ID tracks associated with electron and muon candidates are required to be consistent with
originating from the primary interaction vertex, which is defined as the vertex whose constituent tracks have the highest sum of $p_{\mathrm T}^2$. 
The transverse impact parameter with respect to the beam line, $d_0$, must have a significance satisfying
$|d_0|/\sigma(d_0) < 5\ (3)$ for electrons (muons). For muons, the longitudinal impact parameter, which is the 
distance between the $z$-position of the point of closest approach of the muon track in the ID to the beamline and the $z$-coordinate of the primary vertex, must fulfil $|\Delta z_0| \times\sin\theta < 0.5$~mm.
Both the electrons and muons are required to be isolated with respect to other particles in the event.
The sum of the \pt\ of tracks that fall inside an isolation cone around the lepton (excluding the track of the lepton itself) divided by the lepton \pt\ has to be below a \pt-dependent threshold. The isolation cone size $\Delta R = \sqrt{(\Delta \eta)^2 + (\Delta \phi)^2}$ is defined as $10~\gev$ divided by the lepton \pt and the cone has a maximum size of $\Delta R = 0.3$ for muons and $\Delta R = 0.2$ for electrons. For electrons, calorimeter-based isolation is also required. The sum of the calorimeter transverse energy deposits in an isolation cone of fixed size $\Delta R = 0.2$ (excluding the energy deposits of the electron itself) divided by the electron \pt is used as the discriminating variable. The calorimeter- and track-based isolation criteria depend on \pt\ and $\eta$, and are optimised for an overall efficiency of 98\% (99\%) for electrons (muons). 

The missing transverse momentum is reconstructed as the negative vectorial sum of the calibrated momenta of electrons, muons and jets, where the electrons and muons are required to satisfy the selection criteria described above \cite{etmisspaper}. The jets used in the calculation are reconstructed in the region $|\eta| < 4.9$ from topological clusters \cite{topo1} in the calorimeter using the anti-$k_t$ algorithm \cite{antikt} with a radius parameter of $0.4$. They are calibrated using the method described in Ref.~\cite{topo2} and are required to have $\pt~>~20$~\gev. The computation of \met{} also includes tracks associated with the primary vertex from activity not associated with electrons, muons or jets.

%-------------------------------------------------------------------------------
\section{Event selection and background estimation}
\label{sec:bg}
%-------------------------------------------------------------------------------

Events in the muon channel were recorded by a trigger requiring that at least one muon with $\pt > 50$~\gev\ is found. These muons must be reconstructed in both the MS and the ID. 
In the electron channel during the 2015 (2016) data-taking period, events were recorded by a trigger requiring at least one electron with $\pt > 24~(60)$~\gev\ which satisfied the {\textit{medium}} identification criteria, or at least one electron with $\pt > 120~(140)$~\gev\ which satisfied the {\textit{loose}} identification criteria. The identification criteria for electrons at trigger level are similar to those used in the offline reconstruction~\cite{elepaper}. 

Events recorded by the trigger are further selected by requiring that they contain exactly one lepton. In the muon channel, the magnitude of \met{}\ must exceed $55$~\gev\ and the muon has to fulfil the tight requirements for high-\pt\ muons detailed in Section~\ref{sec:selection} and have $\pt > 55$~\gev. In the electron channel, the electron must satisfy the {\textit{tight}} identification criteria, and the electron \pt\ and the magnitude of \met{} must both be in excess of $65$~\gev. Events in both channels are vetoed if they contain additional leptons
satisfying loosened selection criteria, namely electrons with $\pt > 20$~\gev\ satisfying the {\textit{medium}} identification criteria or muons with $\pt > 20$~\gev\ passing the muon selection without the stringent requirements on the MS track quality. In addition, the transverse mass is required to exceed $110~(130)$~\gev\ in the muon (electron) channel. 
The acceptance times efficiency, defined as the fraction of simulated signal events that pass the event selection described above, is $50\%$ ($47\%$) for the muon channel and $81\%$ ($77\%$) for the electron channel for a \wp\ mass of 2~\tev\ (4~\tev). The difference in lepton sensitivity results from lower muon trigger efficiency and, due to the very strict muon selection criteria applied, a lower muon identification efficiency. 

\begin{figure}
  \centering
  \subfigure[]{\includegraphics[width=0.49\columnwidth]{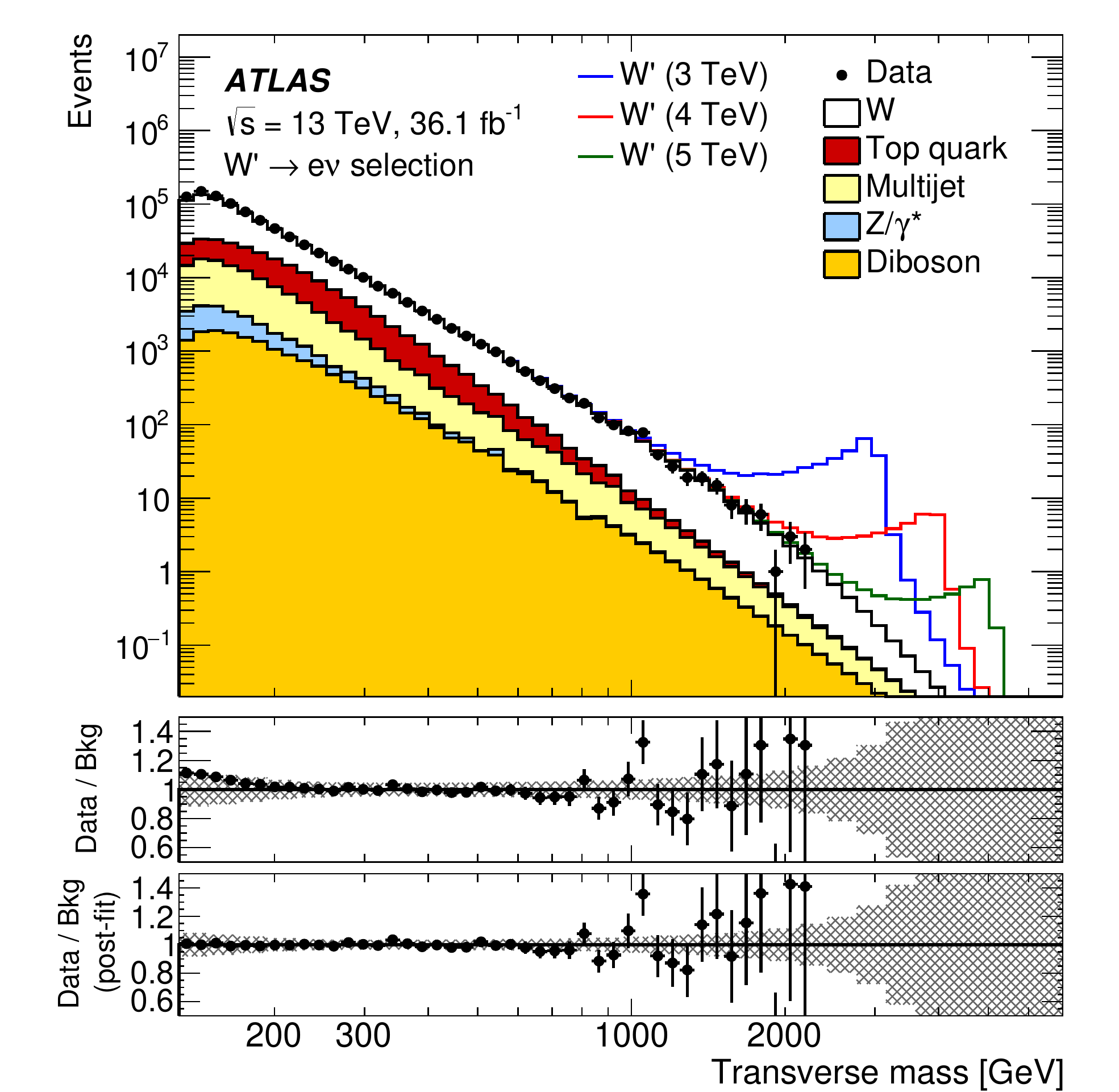} }
  \subfigure[]{\includegraphics[width=0.49\columnwidth]{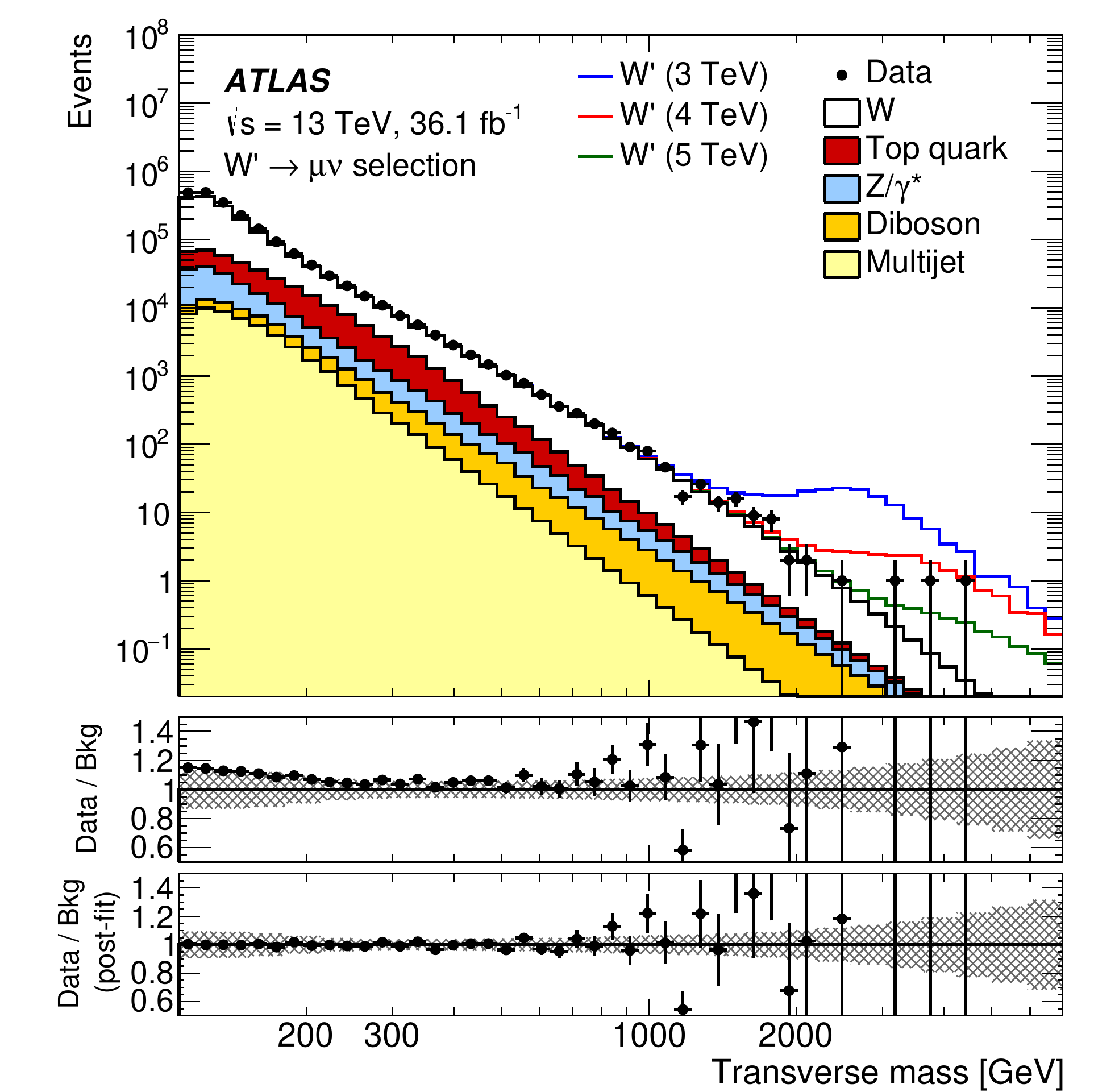} }
  \caption{ 
Transverse mass distributions for events satisfying all selection criteria in the (a) electron and (b) muon channels.
The distributions in data are compared to the stacked sum of all expected backgrounds.
As examples, expected signal distributions for three different SSM \wp\ boson
masses are shown on top of the SM prediction.
The bin width is constant in $\log(m_{\mathrm T})$. 
The middle panels show the ratios of the data to the expected background, with vertical bars representing both data and MC
statistical uncertainties.
The lower panels show the ratios of the data to the adjusted expected background (``post-fit'') that results from the statistical analysis.
The bands in the ratio plots indicate the sum in quadrature of the systematic uncertainties discussed in Section~\ref{sec:systematics},
including the uncertainty in the integrated luminosity.
}
  \label{fig:mT}
\end{figure}
\begin{table*}[htbp]
\fontsize{3mm}{3.5mm}\selectfont
\setlength{\abovecaptionskip}{6pt}
\setlength{\belowcaptionskip}{6pt}
\caption{The numbers of expected events from the total SM background and SSM \wp\ signal and the numbers of observed events in data 
in the electron (top) and muon (bottom) channels
in bins of \mt. The uncertainties given are the combined statistical and systematic uncertainties. 
%Systematic uncertainties are only taken into account if they are estimated to change the total number of expected background events in the respective channel by at least 3\% in at least one bin in $m_T$. 
The systematic uncertainty includes all systematic uncertainties 
%which are included in the statistical analysis 
except the one from the integrated luminosity (3.2\%).
%(see Section~\ref{sec:systematics} for details).
}
\label{tab:backgroundTable}
\begin{tabular}{l r@{ $\pm$ }l r@{ $\pm$ }l r@{ $\pm$ }l r@{ $\pm$ }l r@{ $\pm$ }l r@{ $\pm$ }l r@{ $\pm$ }l }
\multicolumn{15}{c}{Electron channel} \\
\hline
\hline
$m_\mathrm{T}$ [\GeV]\!\!\!\! & \multicolumn{2}{c}{130--200} & \multicolumn{2}{c}{200--400} & \multicolumn{2}{c}{400--600} & \multicolumn{2}{c}{600--1000} & \multicolumn{2}{c}{1000--2000} & \multicolumn{2}{c}{2000--3000} & \multicolumn{2}{c}{3000--7000} \\
\hline
Total SM\!\! &   620\,000 & 70\,000  &   168\,000 & 10\,000 &   9700 & 500  &   2010 & 140  &   232  & 24  &   5.9 & 1.4  &   0.4 & 0.4  \\
\hline
$W^\prime$ (2 \TeV)\!\!\!\!\!\! &   24.3 & 0.9  &  126  & 3     & 199 & 5   &   614 & 14   &   3280 & 50  &   330 & 70  &   0.85 & 0.04 \\
$W^\prime$ (3 \TeV)\!\!\!\!\!\! &   3.83 & 0.08  &  14.2  & 0.2  & 16.1 & 0.4 &   35.7 & 0.4 &   122 & 2  &   229 & 4  &   24 & 5 \\
$W^\prime$ (4 \TeV)\!\!\!\!\!\! &   1.18 & 0.02  &  4.06  & 0.03  & 3.58 & 0.03 &   5.92 & 0.03 &   12.1 & 0.1  &   13.5  & 0.2  &   23.3 & 0.2 \\
$W^\prime$ (5 \TeV)\!\!\!\!\!\! &   0.476 & 0.008  &  1.62  & 0.01  & 1.35 & 0.01 &   1.95 & 0.01 &   2.64 & 0.01  &   1.56 & 0.01  &   3.72 & 0.02  \\
\hline
\hline
Data    &\multicolumn{2}{c}{671\,128}     &\multicolumn{2}{c}{169\,338}     &\multicolumn{2}{c}{9551}      &\multicolumn{2}{c}{1931}       &\multicolumn{2}{c}{246}        &\multicolumn{2}{c}{4} &\multicolumn{2}{c}{0}\\
\hline
\hline
%\end{tabular}
%\end{center}
%muons start
& \multicolumn{2}{c}{} & \multicolumn{2}{c}{} & \multicolumn{2}{c}{} & \multicolumn{2}{c}{} & \multicolumn{2}{c}{} & \multicolumn{2}{c}{} & \multicolumn{2}{c}{} \\ 
%\begin{tabular}{lcccccc}
\multicolumn{15}{c}{Muon channel} \\
\hline
\hline
$m_\mathrm{T}$ [\GeV]\!\!\!\! & \multicolumn{2}{c}{110--200} & \multicolumn{2}{c}{200--400} & \multicolumn{2}{c}{400--600} & \multicolumn{2}{c}{600--1000} & \multicolumn{2}{c}{1000--2000} & \multicolumn{2}{c}{2000--3000} & \multicolumn{2}{c}{3000--7000} \\
\hline
Total SM\!\! &   1\,640\,000 & 200\,000  &   122\,000 & 8000  &   6460 & 330  &   1320 & 90  &   150 & 13  &   4.7 & 0.6  &   0.63 & 0.13  \\ 
\hline 
$W^\prime$ (2 \TeV)\!\!\!\!\!\! &   25.0 & 1.5  &   102 & 6  &   143 & 9       &   420 & 22  &   1720 & 90  &   369 & 28  &   17 & 4  \\ 
$W^\prime$ (3 \TeV)\!\!\!\!\!\! &   3.98 & 0.12  &   10.3 & 0.3  &   10.7 & 0.5  &   26.3 & 1.5  &   84 & 5  &   98 & 6  &   39.3 & 3.4  \\ 
$W^\prime$ (4 \TeV)\!\!\!\!\!\! &   1.20 & 0.03  &   2.80 & 0.07  &   2.36 & 0.09  &   4.07 & 0.19  &   8.1 & 0.5  &   8.8 & 0.6  &   11.1 & 0.9  \\ 
$W^\prime$ (5 \TeV)\!\!\!\!\!\! &   0.485 & 0.012  &   1.12 & 0.03  &   0.88 & 0.03  &   1.27 & 0.05  &   1.7 & 0.1  &   0.99 & 0.07  &   1.7 & 0.1  \\ 
\hline
\hline
Data    &\multicolumn{2}{c}{1\,862\,326}     &\multicolumn{2}{c}{128\,155}      &\multicolumn{2}{c}{6772}      &\multicolumn{2}{c}{1392}        &\multicolumn{2}{c}{177}       &\multicolumn{2}{c}{3} &\multicolumn{2}{c}{3}\\
%Data & 341220 & 80651 & 30485 & 1612 & 314 & 40 & 0 \\
\hline
\hline
\end{tabular}
\end{table*}

The expected number of background events is calculated as the sum of the data-driven and simulated background estimates described in Section~\ref{sec:MC}. Figure~\ref{fig:mT} displays the \mt\ distribution in the electron and muon channels. The expected and observed number of events for some wider \mt\ ranges are shown also in Table~\ref{tab:backgroundTable}.
For all values of \mt, the background is dominated by $W \rightarrow \ell \nu$ production, which constitutes about 85\% of the total background at \mt\ $>1$~\TeV. As examples, Figure~\ref{fig:mT} also shows the expected signal distributions for a few \wpssm\ boson masses on top of the SM prediction. The effect of the momentum resolution is clearly visible when comparing the shapes of the three reconstructed \wpssm\ signals in the electron and muon channels. 
The middle panels of Figure~\ref{fig:mT} show the ratio of the data to the SM predictions. 
The data are systematically above the predicted background at low \mt, but still within the total systematic uncertainty, which is dominated by the \met{}-related systematic uncertainties in this region. 
The bottom panels of Figure~\ref{fig:mT} show the ratio of the data to 
the adjusted background that results from a common fit to the electron and muon channels within the statistical analysis described in Section~\ref{sec:result}.
This ratio agrees well with unity.

%-------------------------------------------------------------------------------
\section{Systematic uncertainties}
\label{sec:systematics}
\begin{table}
\caption{Systematic uncertainties in the expected number of events as estimated for the total background and for signal
with a \wpssm\ mass of 2~(4)~\TeV. The uncertainty is estimated with the binning shown in Figure~\ref{fig:mT} at
\mt\ $=$ 2 (4)~\TeV\ for the background and in a three-bin window around \mt\ $=$ 2 (4)~\TeV\ for the signal.
Uncertainties that are not applicable are denoted ``{\textsc {n/a}}'', and ``negl.'' means that the uncertainty is not included in the statistical analysis. 
Sources of uncertainties not included in the table are neglected in the statistical analysis. \label{tab:syst}}
\begin{center}
\centering
\small
\begin{tabular}{l|cc|cc}
\toprule
\toprule
Source  &  \multicolumn{2}{c|}{Electron channel}  &  \multicolumn{2}{c}{Muon channel}  \\
&  Background  &  Signal  &  Background  &  Signal  \\
\midrule
Trigger &negl. (negl.) & negl. (negl.) &\syspair{2}{2} & \syspair{2}{2}\\
Lepton reconstruction &\multirow{2}{*}{negl. (negl.)} & \multirow{2}{*}{negl. (negl.)} &\multirow{2}{*}{\syspair{5}{6}} & \multirow{2}{*}{\syspair{5}{7}}\\
and identification & & & & \\
Lepton momentum &\multirow{2}{*}{\syspair{3}{3}} & \multirow{2}{*}{\syspair{4}{3}} &\multirow{2}{*}{\syspair{3}{9}} & \multirow{2}{*}{\syspair{1}{1}}\\
scale and resolution & & & & \\
\met{} resolution and scale &\syspair{<0.5}{<0.5} &\syspair{<0.5}{<0.5} &\syspair{<0.5}{1} &\syspair{1}{1}\\
Jet energy resolution &\syspair{<0.5}{<0.5} &\syspair{<0.5}{<0.5} &\syspair{<0.5}{<0.5} &\syspair{<0.5}{<0.5}\\
Pile-up &\syspair{1}{<0.5} &\syspair{1}{<0.5} &\syspair{<0.5}{1} &\syspair{1}{<0.5}\\
\midrule
Multijet background &\syspair{7}{70} & {\textsc {n/a}} ({\textsc {n/a}}) & \syspair{1}{1} & {\textsc {n/a}} ({\textsc {n/a}})\\
Top extrapolation &\syspair{1}{1} & {\textsc {n/a}} ({\textsc {n/a}}) &\syspair{4}{8} & {\textsc {n/a}} ({\textsc {n/a}})\\
Diboson extrapolation &\syspair{4}{20} & {\textsc {n/a}} ({\textsc {n/a}}) &\syspair{4}{10} & {\textsc {n/a}} ({\textsc {n/a}})\\
PDF choice for DY &\syspair{1}{13} & {\textsc {n/a}} ({\textsc {n/a}}) &\syspair{<0.5}{1} & {\textsc {n/a}} ({\textsc {n/a}})\\
PDF variation for DY &\syspair{8}{15} & {\textsc {n/a}} ({\textsc {n/a}}) &\syspair{7}{11} & {\textsc {n/a}} ({\textsc {n/a}})\\
EW corrections for DY &\syspair{4}{7} & {\textsc {n/a}} ({\textsc {n/a}}) &\syspair{4}{5} & {\textsc {n/a}} ({\textsc {n/a}})\\
\midrule
Luminosity &\syspair{3}{3} &\syspair{3}{3} &\syspair{3}{3} &\syspair{3}{3}\\
\midrule
Total &\syspair{13}{76} & \syspair{5}{5} &\syspair{12}{21} & \syspair{6}{8}\\
\bottomrule
\bottomrule
\end{tabular}
\end{center}
\end{table}

The systematic uncertainties arise from experimental and theoretical sources. They are summarised in Table~\ref{tab:syst} and described in the following subsections.

\subsection{Uncertainties from the reconstruction of electrons, muons, and \met }
Experimental systematic uncertainties arise from the trigger, reconstruction, identification and isolation efficiencies for leptons \cite{elepaper, muonpubnote}, and the calculation of the missing transverse momentum~\cite{etmisspaper}. They include also the effects of the energy and momentum scale and resolution uncertainties \cite{elecalibpaper, muonpubnote, topo2}. 

The electron and muon offline reconstruction, identification and isolation efficiencies, and their respective uncertainties, are assessed up to $\pt \approx 100$~\gev\ using leptonic decays of $Z$ boson candidates found in data. The ratio of the efficiency measured in data to that of the MC simulation is then used to correct the MC prediction \cite{elepaper,muonpubnote}.  
For higher-\pt\ electrons, an additional uncertainty of 1.5\% is estimated for the {\textit{tight}} identification working point. This uncertainty is based on the differences observed in the electron shower shapes in the EM calorimeters between data and MC simulation around the $Z \to ee$ mass peak, which are propagated to the high-\et\ electron sample.
For the isolation efficiencies, an uncertainty of $2$\% and $5$\% is estimated for $150 < \pt < 500$ \gev\ and above $500$~\gev, respectively, using $Z/\gamma^*$ candidates in data. 
For the identification of high-\pt muons, the uncertainty is determined conservatively from simulation studies and amounts to 2--3\% per \tev.
For the isolation criterion, the uncertainty associated with the extrapolation to high-\pt\ muons is estimated to be 1\%. Systematic uncertainties related to the electron trigger are negligible. For the muon trigger the systematic uncertainty is estimated using the same methodology as in Ref.~\cite{muontrigger}, which results in an overall uncertainty of about 2\%.

The main systematic uncertainties in \met{}\ arise from the jet energy resolution uncertainties \cite{topo2} and the contribution from tracks originating from the primary vertex and arising from activity not associated with electrons, muons or jets \cite{etmisspaper}.
The uncertainties due to the jet energy and \met{}\ resolutions are small at large \mt, while they are the dominant contributions
to the total uncertainty at small \mt. The jet energy scale uncertainties are found to be negligible.

\subsection{Theoretical uncertainties}
\label{sec:sysTheory}

Theoretical uncertainties are related to the production cross-sections estimated from MC simulation. The effects when propagated
to the total background estimate are significant for $W$ and $Z/\gamma^*$ production, but 
negligible for top-quark and diboson production. No theoretical uncertainties are considered for the \wp\ boson signal in the statistical analysis. 

Theoretical uncertainties in the $W$ and $Z/\gamma^*$ background prediction arise from the PDF uncertainties, the value of the strong coupling constant $\alpha_{\mathrm s}$, and higher-order corrections. 
The dominant effect comes from the PDF uncertainty, which is obtained from the 90\%~CL CT14NNLO PDF uncertainty set using \vrap\ to calculate the NNLO cross-section as a function of the boson mass. Rather than using the original 28 CT14 uncertainty eigenvectors, a re-diagonalised set of seven PDF eigenvectors, as provided by the authors of the CT14 PDF using MP4LHC~\cite{gao:2013bia, butterworth:2015oua}, is used. The cross-section variation associated with each of these eigenvectors has a characteristic mass dependence and the sum in quadrature of these eigenvector variations matches the original CT14NNLO uncertainty envelope well. 
This sum is shown as ``PDF variation'' in Table~\ref{tab:syst}. An additional uncertainty is 
derived to account for the choice of the nominal PDF set used.
The central values of the CT14NNLO PDF set are compared to the MMHT2014~\cite{Harland-Lang:2014zoa} and NNPDF3.0~\cite{Ball:2014uwa} PDF sets. A comparison between these PDF sets shows that the central value for NNPDF3.0 fall outside the ``PDF variation'' uncertainty at large \mt. Thus, an envelope of the 
``PDF variation'' and the NNPDF3.0 central value is formed, where 
the former is subtracted in quadrature from this envelope, 
and the remaining part, which is only non-zero when the NNPDF3.0 central value is outside the ``PDF variation'' uncertainty, is quoted as ``PDF choice''.
The PDF uncertainties are the same at the generator level for the electron and muon channels, but result in different uncertainties at reconstruction level. 
The uncertainty is larger in the electron channel due to the better energy resolution: there is less migration of events with low generator-level invariant mass, where the PDF uncertainty is smaller, into the high-\mt\ region in this channel. 

Uncertainties in the electroweak corrections are determined as the difference between the additive approach ($1+\delta_{\mathrm{EW}} + \delta_{\mathrm{QCD}}$) and a
factorised approach ($(1+\delta_{\mathrm{EW}}) \times (1+\delta_{\mathrm{QCD}})$) for the EW corrections in the combination of higher-order EW ($\delta_{\mathrm{EW}}$) and QCD ($\delta_{\mathrm{QCD}}$) effects. 
Uncertainties due to higher-order QCD corrections on the $Z/\gamma^*$ process are estimated by varying the renormalisation and factorisation scales 
simultaneously up or down by a factor of two. The uncertainty due to $\alpha_{\mathrm s}$ is assessed by changing the value of $\alpha_{\mathrm s}$ by as much as 0.003 
from the nominal value $\alpha_{\mathrm s}(m_Z) = 0.118$ used by the CT14NNLO PDF set. The uncertainties from the scales and $\alpha_{\mathrm s}$ are both found to be negligible.

Theoretical uncertainties are also considered for the top-quark and diboson backgrounds. An uncertainty in the $t\bar{t}$ cross-section of $^{+20}_{-29}$~pb  arises from the independent variation of the factorisation and renormalisation scales, while an uncertainty of $\pm 35$~pb is associated with variations in the PDF and $\alpha_{\mathrm s}$, following 
the PDF4LHC prescription (see Ref.~\cite{Botje:2011sn} and references therein) with the MSTW2008 68\% CL NNLO~\cite{Martin:2009bu}, 
CT10 NNLO~\cite{Gao:2013xoa} and NNPDF2.3 NNLO~\cite{Ball:2012cx} PDF sets.
As this background constitutes only a small fraction of the overall background, these normalisation uncertainties are negligible.
Furthermore, the modelling of the top-quark background is found to be adequate in a data control region defined by requiring the presence of an additional muon (electron) in events passing the electron (muon) selection. For the diboson background, the theoretical normalisation uncertainty is conservatively
taken to be 30\%, and this has a negligible effect due to the small contribution of this background.

\subsection{Background modelling uncertainties}
The dominant systematic uncertainties in the multijet, top-quark and diboson backgrounds at high \mt\ are due to 
the extrapolations. These uncertainties are evaluated by varying both the functional form of the fit functions and the fit range as detailed in Section \ref{sec:MC}. The envelope of all variations is assigned for the uncertainty. This results in the largest source of background-related systematic uncertainty at large \mt\ values in this analysis.

The multijet background uncertainty in the electron (muon) channel includes a 15\% (100\%) normalisation uncertainty. This uncertainty is dominated by the dependence of the factor $f$ (see Section \ref{sec:MC}) on the selection requirements used for the background-enriched sample definition.

For the \mt\ region below 700--800 \gev, for which there are not many more MC events than data events,
the MC statistical uncertainty is accounted for in the analysis. 

The modelling of the pile-up especially affects the calculation of \met{}. 
A pile-up modelling uncertainty is estimated by varying the distribution of pile-up events in the reweighting of the MC, as detailed in Section~\ref{sec:MC}, to cover the uncertainty on the ratio between the predicted and measured inelastic cross-sections~\cite{Aaboud:2016mmw}.

\subsection{Luminosity}
The uncertainty in the combined 2015 and 2016 integrated luminosity is 3.2\%. Following a methodology similar to that detailed in Ref.~\cite{lumipaper}, it is derived from a preliminary calibration of the luminosity scale using $x$--$y$ beam-separation scans performed in August 2015 and May 2016. 

%-------------------------------------------------------------------------------

%-------------------------------------------------------------------------------
\section{Results}
\label{sec:result}
%-------------------------------------------------------------------------------

For the statistical analysis of the results presented in this section, the same methodology is applied as in the previous ATLAS \wp\ search~\cite{atlas_13tev_pub} and is described briefly here.
The compatibility between the data and the predicted background is evaluated with a profile-likelihood ratio test quantifying the probability that the background 
fluctuates to give a signal-like excess equal to or larger than what is observed. 
The likelihood functions in the ratio are products of Poisson probabilities
over all bins in the transverse mass distribution (as shown in Figure~\ref{fig:mT})
and log-normal constraints for the variations in signal and background yields associated with systematic uncertainties.
In the denominator of the likelihood ratio, the likelihood function is maximised assuming the presence of a signal above the expected background, and in the numerator assuming the background-only hypothesis. 
To model the signal, \wpssm\ templates binned in \mt\ 
are used for a series 
of \wpssm\ masses in the search range $150~\gev\ \leq$ \mwp\ $\leq 6000~\gev$. Figure~\ref{fig:mT} displays a few examples of these templates. 
No significant excesses are observed in the data. The most significant excess is at $m_{\wp} = 350$~\gev\ in the electron channel, with a local significance of $2.0 \sigma$. In the muon channel, the most significant excess is at high mass, with a maximum local significance of $1.8 \sigma$ at $m_{\wp} \approx 5$~\tev. These excesses correspond to a global significance of $0.1 \sigma$ in each channel when the look-elsewhere effect~\cite{LLE} is taken into account.

Based on the above findings, upper limits on the cross-section for producing a \wpssm\ boson times its branching ratio to only one lepton generation (\xbr) are computed at the 95\% CL as a function of the \wpssm\ boson mass. The limits are calculated in a Bayesian analysis~\cite{bayes} with a uniform positive prior probability distribution for \xbr. 
The observed upper limits are extracted by comparing 
data to the expected background and signal using \wpssm\ templates for 
the same range of signal masses as for the profile-likelihood ratio test. The expected limits are derived from pseudo-experiments obtained from the estimated background distributions. The median of the distribution of the limits from the pseudo-experiments is taken as the expected limit, and $1 \sigma$ and $2 \sigma$ bands are defined as the ranges containing 
respectively 68\% and 95\% of the limits obtained with the pseudo-experiments.

\begin{figure}[htbp]
\centering
\subfigure[]{\includegraphics[width=0.51\columnwidth]{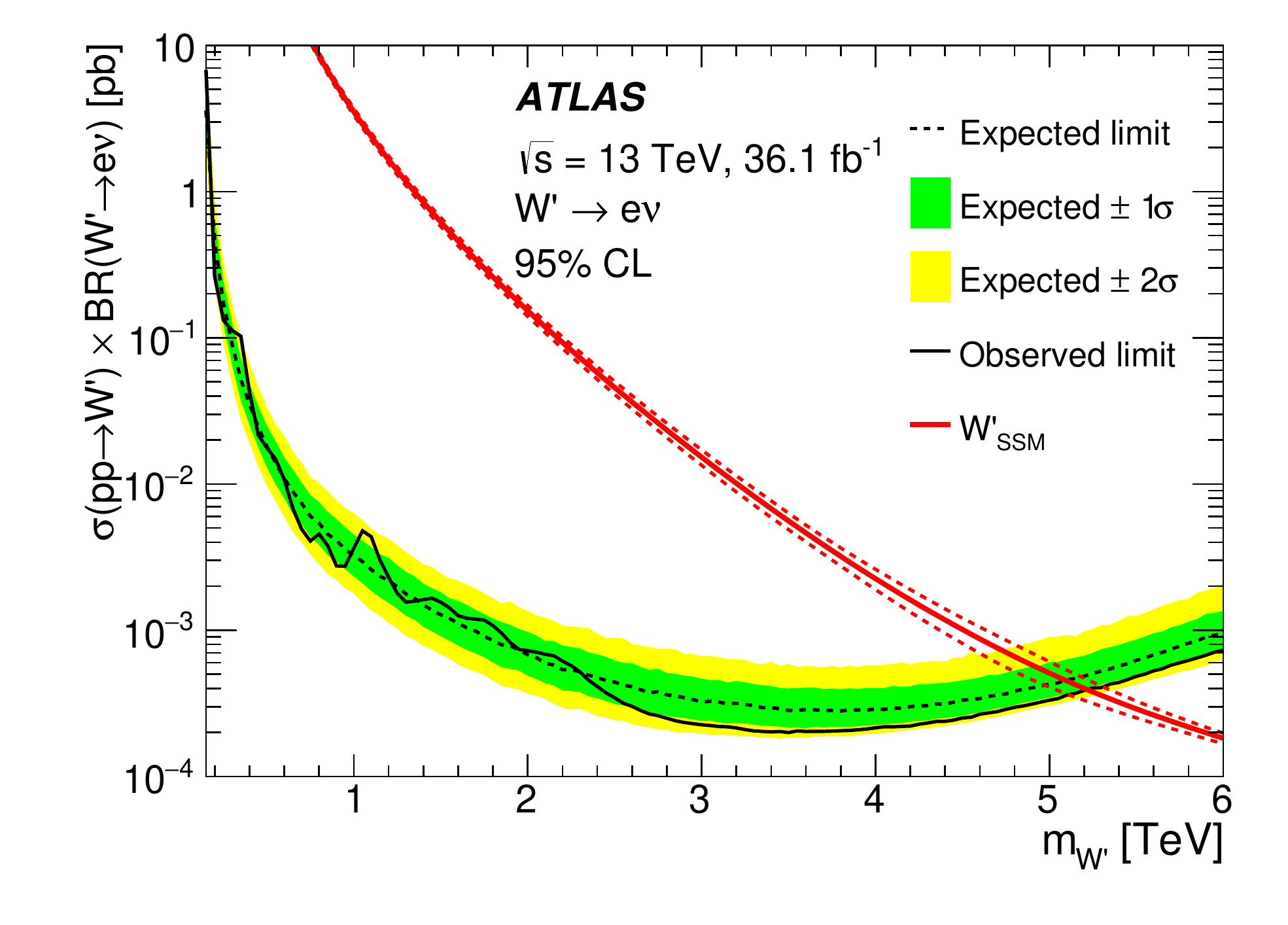} }
\subfigure[]{\includegraphics[width=0.51\columnwidth]{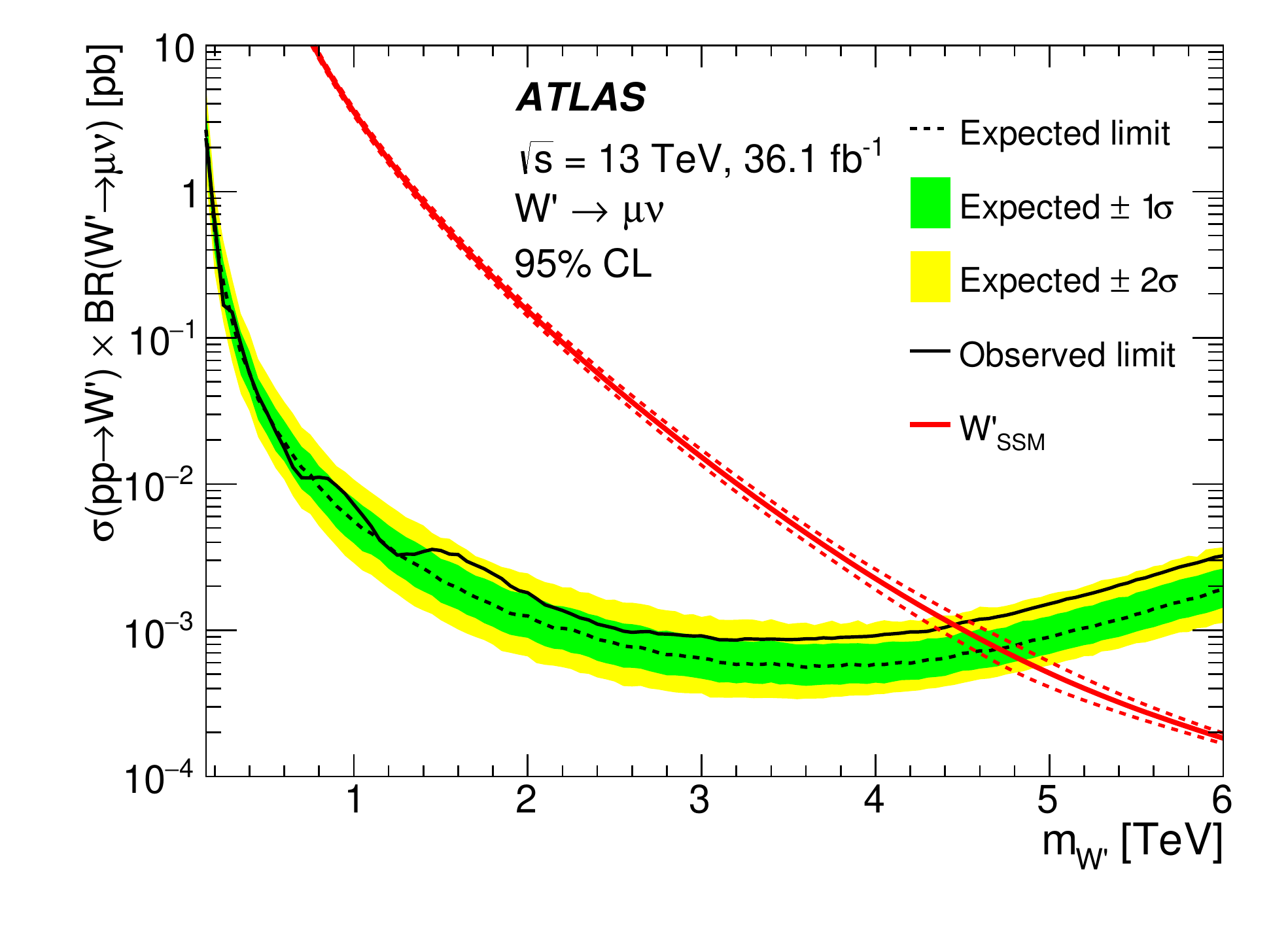} }
\subfigure[]{\includegraphics[width=0.51\columnwidth]{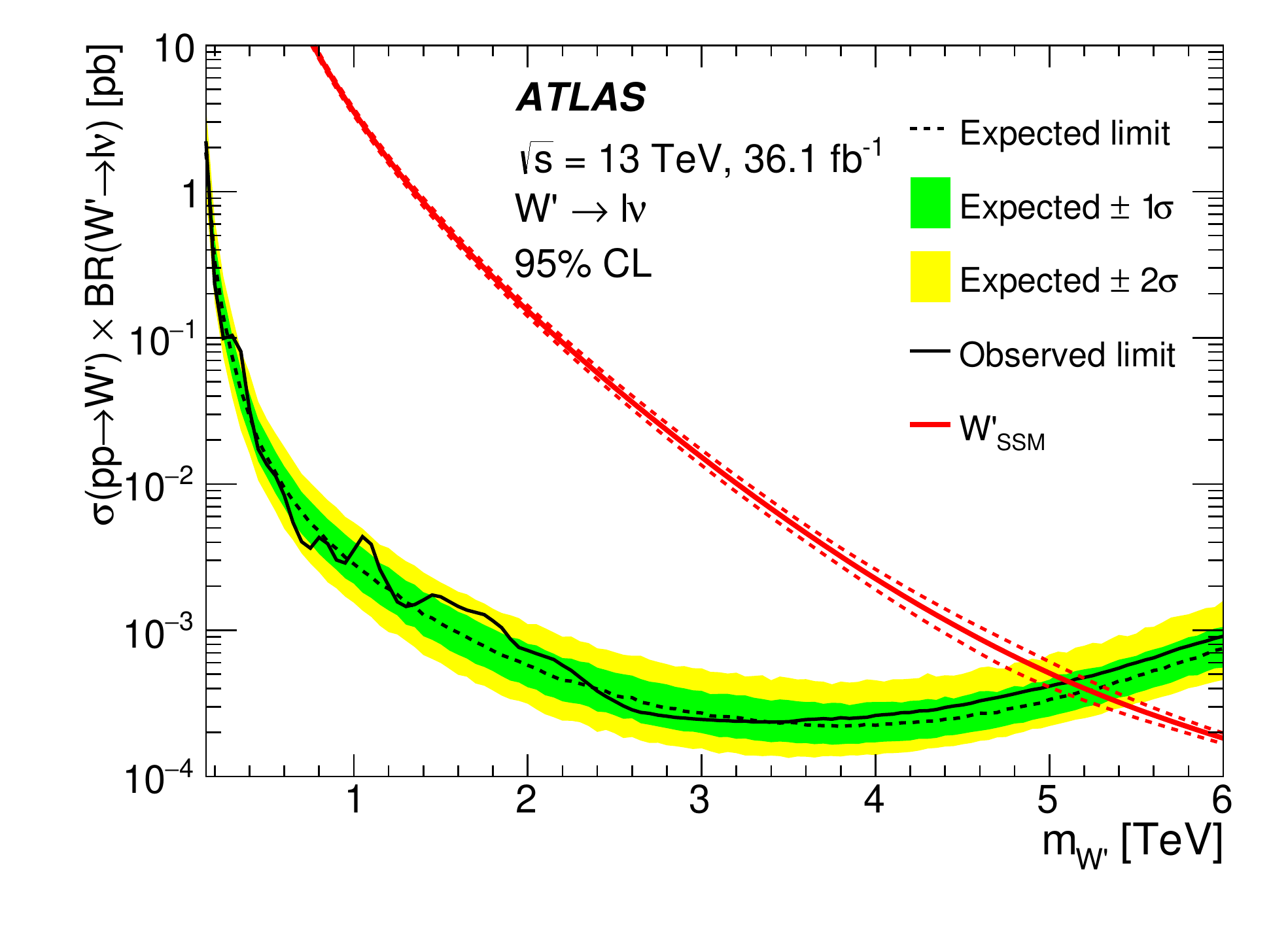} }
  \caption{
Observed (solid black line) and expected (dashed black line) upper limits on cross-section times branching ratio (\xbr) as a function of the SSM \wp\ boson mass in the (a) electron, (b) muon and (c) combined electron and muon channels. The $1\sigma$  (green) and $2\sigma$ (yellow) expected limit bands are also shown.
The predicted \xbr\ for SSM \wp\ production is shown as a red solid line. 
For illustration the uncertainties in \xbr\ from the PDF, $\alpha_{\mathrm s}$ and the renormalisation and factorisation scales are also shown as red-dashed lines.
}
  \label{fig:e_mu_limits}
\end{figure}
The 95\% CL upper limits on \xbr\ as a function of the \wpssm\ mass are shown in Figure~\ref{fig:e_mu_limits} separately for the electron and muon channels and for the combination of the two channels. The theoretical uncertainties and the uncertainties in \met{}, jet energy resolution and luminosity are treated as correlated between the channels. 
The expected upper limit on \xbr\ is stronger in the electron channel. This results from the larger acceptance times efficiency and the better momentum resolution (see Section \ref{sec:bg}). 
Figure~\ref{fig:e_mu_limits} also shows the predicted \xbr\ for the \wpssm\ boson as a function of its mass as well as the uncertainties from the PDF, $\alpha_{\mathrm s}$ and the factorisation and renormalisation scales
derived using the same prescription as used for the $W$ boson production. The observed (expected) lower mass limit for a \wpssm\ boson, as summarised in Table~\ref{tab:limits_mass_wp}, is $5.1~(5.2)$~\tev\
 for the combination of the electron and muon channels.
 This corresponds to an improvement of approximately 1~\tev\ in mass reach compared to the previous ATLAS analysis \cite{atlas_13tev_pub}, which was based on a subset of the data used in this analysis.
% limits_mass_wp.tex
\begin{table}[!htbp]
  \caption{Expected and observed 95\% CL lower limit on the \wpssm\ mass in the electron and muon channels and their combination.}
  \centering
  \begin{tabular}{c|cc}
    \hline
    \hline
    &  \multicolumn{2}{c}{\mwp\ lower limit [\TeV]} \\
    Decay     &  Expected & Observed \\
    \hline
    \wpe  & 5.1 & 5.2 \\
    \wpmu & 4.7 & 4.5 \\ \hline
    \wpl  & 5.2 & 5.1 \\
    \hline
    \hline
  \end{tabular}
  \label{tab:limits_mass_wp}
\end{table}

%

% All figures and tables should appear before the summary and conclusion.
% The package placeins provides the macro \FloatBarrier to achieve this.
\FloatBarrier

%-------------------------------------------------------------------------------
\section{Conclusion}
\label{sec:conclusion}
The results of a search for a new heavy gauge boson decaying to final states with a high-\pt\ electron or muon and large missing transverse momentum are reported. The analysis uses 36.1 \ifb\ of $\sqrt{s} = 13$~\tev\ $pp$ collision data recorded by the ATLAS detector at the Large Hadron Collider in 2015 and 2016. Examining the transverse mass spectrum, no significant excess above the expected Standard Model background is observed. Exclusion limits at 95\% CL are placed on the mass of benchmark Sequential Standard Model \wp\ bosons. Masses for \wpssm\ bosons up to 5.1~\tev\ are excluded by the combination of the electron and muon channels.
This exceeds the previous limit from ATLAS, derived from a similar analysis based on 3.2 \ifb\ of $\sqrt{s} = 13$~\tev\ data, by 1 \tev.

%-------------------------------------------------------------------------------

%-------------------------------------------------------------------------------
\section*{Acknowledgements}
%-------------------------------------------------------------------------------

% Acknowledgements for papers with collision data
% Version 06-Mar-2017

% Standard acknowledgements start here
%----------------------------------------------
We thank CERN for the very successful operation of the LHC, as well as the
support staff from our institutions without whom ATLAS could not be
operated efficiently.

We acknowledge the support of ANPCyT, Argentina; YerPhI, Armenia; ARC, Australia; BMWFW and FWF, Austria; ANAS, Azerbaijan; SSTC, Belarus; CNPq and FAPESP, Brazil; NSERC, NRC and CFI, Canada; CERN; CONICYT, Chile; CAS, MOST and NSFC, China; COLCIENCIAS, Colombia; MSMT CR, MPO CR and VSC CR, Czech Republic; DNRF and DNSRC, Denmark; IN2P3-CNRS, CEA-DSM/IRFU, France; SRNSF, Georgia; BMBF, HGF, and MPG, Germany; GSRT, Greece; RGC, Hong Kong SAR, China; ISF, I-CORE and Benoziyo Center, Israel; INFN, Italy; MEXT and JSPS, Japan; CNRST, Morocco; NWO, Netherlands; RCN, Norway; MNiSW and NCN, Poland; FCT, Portugal; MNE/IFA, Romania; MES of Russia and NRC KI, Russian Federation; JINR; MESTD, Serbia; MSSR, Slovakia; ARRS and MIZ\v{S}, Slovenia; DST/NRF, South Africa; MINECO, Spain; SRC and Wallenberg Foundation, Sweden; SERI, SNSF and Cantons of Bern and Geneva, Switzerland; MOST, Taiwan; TAEK, Turkey; STFC, United Kingdom; DOE and NSF, United States of America. In addition, individual groups and members have received support from BCKDF, the Canada Council, CANARIE, CRC, Compute Canada, FQRNT, and the Ontario Innovation Trust, Canada; EPLANET, ERC, ERDF, FP7, Horizon 2020 and Marie Sk{\l}odowska-Curie Actions, European Union; Investissements d'Avenir Labex and Idex, ANR, R{\'e}gion Auvergne and Fondation Partager le Savoir, France; DFG and AvH Foundation, Germany; Herakleitos, Thales and Aristeia programmes co-financed by EU-ESF and the Greek NSRF; BSF, GIF and Minerva, Israel; BRF, Norway; CERCA Programme Generalitat de Catalunya, Generalitat Valenciana, Spain; the Royal Society and Leverhulme Trust, United Kingdom.

The crucial computing support from all WLCG partners is acknowledged gratefully, in particular from CERN, the ATLAS Tier-1 facilities at TRIUMF (Canada), NDGF (Denmark, Norway, Sweden), CC-IN2P3 (France), KIT/GridKA (Germany), INFN-CNAF (Italy), NL-T1 (Netherlands), PIC (Spain), ASGC (Taiwan), RAL (UK) and BNL (USA), the Tier-2 facilities worldwide and large non-WLCG resource providers. Major contributors of computing resources are listed in Ref.~\cite{ATL-GEN-PUB-2016-002}.
%----------------------------------------------

%\input{acknowledgements/Acknowledgements}

%The \texttt{atlaslatex} package contains the acknowledgements that were valid 
%at the time of the release you are using.
%These can be found in the \texttt{acknowledgements} subdirectory.
%When your ATLAS paper or PUB/CONF note is ready to be published,
%download the latest set of acknowledgements from:\\
%\url{https://twiki.cern.ch/twiki/bin/view/AtlasProtected/PubComAcknowledgements}

%The supporting notes for the analysis should also contain a list of contributors.
%This information should usually be included in \texttt{mydocument-metadata.tex}.
%The list should be printed either here or before the table of contents.

%-------------------------------------------------------------------------------
%\clearpage
%\appendix
%\part*{Appendix}
%\addcontentsline{toc}{part}{Appendix}
%-------------------------------------------------------------------------------

%In a paper, an appendix is used for technical details that would otherwise disturb the flow of the paper.
%Such an appendix should be printed before the Bibliography.

%-------------------------------------------------------------------------------
% If you use biblatex and either biber or bibtex to process the bibliography
% just say \printbibliography here
\printbibliography
% If you want to use the traditional BibTeX you need to use the syntax below.
%\bibliographystyle{bibtex/bst/atlasBibStyleWoTitle}
%\bibliography{mydocument,bibtex/bib/ATLAS}
%-------------------------------------------------------------------------------

%-------------------------------------------------------------------------------
% Print the list of contributors to the analysis
% The argument gives the fraction of the text width used for the names
%-------------------------------------------------------------------------------
%\clearpage
%\PrintAtlasContribute{0.30}
\newpage 
% ATLAS Collaboration author list
% Data extracted on 01-May-2017 for paper reference EXOT-2016-06
% \documentclass[11pt]{article}
% \usepackage{a4wide}\begin{document}
\begin{flushleft}
{\Large The ATLAS Collaboration}

\bigskip

M.~Aaboud$^\textrm{\scriptsize 137d}$,
G.~Aad$^\textrm{\scriptsize 88}$,
B.~Abbott$^\textrm{\scriptsize 115}$,
O.~Abdinov$^\textrm{\scriptsize 12}$$^{,*}$,
B.~Abeloos$^\textrm{\scriptsize 119}$,
S.H.~Abidi$^\textrm{\scriptsize 161}$,
O.S.~AbouZeid$^\textrm{\scriptsize 139}$,
N.L.~Abraham$^\textrm{\scriptsize 151}$,
H.~Abramowicz$^\textrm{\scriptsize 155}$,
H.~Abreu$^\textrm{\scriptsize 154}$
% ...
\bigskip
\\
% ...
$^{12}$ Institute of Physics, Azerbaijan Academy of Sciences, Baku, Azerbaijan\\
% ...
$^{88}$ CPPM, Aix-Marseille Universit{\'e} and CNRS/IN2P3, Marseille, France\\
% ...
$^{115}$ Homer L. Dodge Department of Physics and Astronomy, University of Oklahoma, Norman OK, United States of America\\
% ...
$^{119}$ LAL, Univ. Paris-Sud, CNRS/IN2P3, Universit{\'e} Paris-Saclay, Orsay, France\\
% ...
$^{137}$ $^{(a)}$ Facult{\'e} des Sciences Ain Chock, R{\'e}seau Universitaire de Physique des Hautes Energies - Universit{\'e} Hassan II, Casablanca; $^{(b)}$ Centre National de l'Energie des Sciences Techniques Nucleaires, Rabat; $^{(c)}$ Facult{\'e} des Sciences Semlalia, Universit{\'e} Cadi Ayyad, LPHEA-Marrakech; $^{(d)}$ Facult{\'e} des Sciences, Universit{\'e} Mohamed Premier and LPTPM, Oujda; $^{(e)}$ Facult{\'e} des sciences, Universit{\'e} Mohammed V, Rabat, Morocco\\
% ...
$^{139}$ Santa Cruz Institute for Particle Physics, University of California Santa Cruz, Santa Cruz CA, United States of America\\
% ...
$^{151}$ Department of Physics and Astronomy, University of Sussex, Brighton, United Kingdom\\
% ...
$^{154}$ Department of Physics, Technion: Israel Institute of Technology, Haifa, Israel\\
$^{155}$ Raymond and Beverly Sackler School of Physics and Astronomy, Tel Aviv University, Tel Aviv, Israel\\
% ...
$^{161}$ Department of Physics, University of Toronto, Toronto ON, Canada\\
% ...
$^{a}$ Also at Department of Physics, King's College London, London, United Kingdom\\
$^{b}$ Also at Institute of Physics, Azerbaijan Academy of Sciences, Baku, Azerbaijan\\
$^{c}$ Also at Novosibirsk State University, Novosibirsk, Russia\\
$^{d}$ Also at TRIUMF, Vancouver BC, Canada\\
$^{e}$ Also at Department of Physics {\&} Astronomy, University of Louisville, Louisville, KY, United States of America\\
% ...
$^{*}$ Deceased
\end{flushleft}

% \end{document}
% Created with xml2latex.py

%-------------------------------------------------------------------------------
%\clearpage
%\appendix
%\part*{Auxiliary material}
%\addcontentsline{toc}{part}{Auxiliary material}
%
%\input{Auxiliary}
%-------------------------------------------------------------------------------

%In an ATLAS paper, auxiliary plots and tables that are supposed to be made public 
%should be collected in an appendix that has the title \enquote{Auxiliary material}.
%This appendix should be printed after the Bibliography.
%At the end of the paper approval procedure, this information can be split into a separate document
%-- see \texttt{atlas-auxmat.tex}.

%In an ATLAS note, use the appendices to include all the technical details of your work
%that are relevant for the ATLAS Collaboration only (e.g.\ dataset details, software release used).
%This information should be printed after the Bibliography.

\end{document}